\documentclass[10pt]{article}
\usepackage{geometry}                
\usepackage[parfill]{parskip}    
\usepackage{graphicx}
\usepackage{amsmath, amsthm, amsfonts}
\usepackage{lipsum}
\usepackage{newtxtext,newtxmath}

\newcommand{\lpp}[1]{\hat{\Lambda}^+_{#1}}
\newcommand{\lm}[1]{\hat{\Lambda}^-_{#1}}
\newcommand{\lpm}[1]{\hat{\Lambda}^\pm_{#1}}

\newcommand{\dpp}[1]{\hat\Sigma^+_{#1}}
\newcommand{\dm}[1]{\hat\Sigma^-_{#1}}
\newcommand{\dpm}[1]{\hat\Sigma^\pm_{#1}}

\newcommand{\Spp}[1]{\hat{\mathcal{S}}^+_{#1}}
\newcommand{\Sm}[1]{\hat{\mathcal{S}}^-_{#1}}
\newcommand{\Spm}[1]{\hat{\mathcal{S}}^\pm_{#1}}

\newcommand{\Lpp}[1]{\hat{\mathcal{L}}^+_{#1}}
\newcommand{\Lm}[1]{\hat{\mathcal{L}}^-_{#1}}
\newcommand{\Lpm}[1]{\hat{\mathcal{L}}^\pm_{#1}}

\newcommand{\lpc}[1]{ \Lambda^+_{#1}}
\newcommand{\lmc}[1]{\Lambda^-_{#1}}
\newcommand{\lpmc}[1]{\Lambda^\pm_{#1}}

\newcommand{\spc}[1]{\Sigma^+_{#1}}
\newcommand{\smc}[1]{\Sigma^-_{#1}}
\newcommand{\spmc}[1]{\Sigma^\pm_{#1}}

\newcommand{\sipc}[1]{\mathcal{S}^s_{#1}}
\newcommand{\anpc}[1]{\mathcal{S}^a_{#1}}

\newcommand{\Spc}[1]{\mathcal{S}^+_{#1}}
\newcommand{\Smc}[1]{ \mathcal{S}^-_{#1}}
\newcommand{\Spmc}[1]{\mathcal{S}^\pm_{#1}}

\newcommand{\Lpc}[1]{\mathcal{L}^+_{#1}}
\newcommand{\Lmc}[1]{\mathcal{L}^-_{#1}}
\newcommand{\Lpmc}[1]{\mathcal{L}^\pm_{#1}}

\newcommand{\sect}[1]{\setcounter{equation}{0}\section{#1}}

\usepackage{color}
     \usepackage[colorlinks]{hyperref}
\hypersetup{linkcolor=blue,%
citecolor=red,%
urlcolor=cyan}
\usepackage{url} 
\usepackage{epstopdf}
\title{Quantum, classical symmetries and action-angle variables   
 \\
by factorization of superintegrable systems 
}

\author{ 
\c{S}eng\"ul Kuru$^a$
\footnote{kuru@science.ankara.edu.tr, ORCID: 
\href{https://orcid.org/0000-0001-6380-280X}{0000-0001-6380-280X}}\,,
Javier Negro$^b$
\footnote{jnegro@fta.uva.es, ORCID: 
\href{http://orcid.org/0000-0002-0847-6420}{0000-0002-0847-6420}}\,,
 Sergio Salamanca$^b$
\footnote{sergio.salamanca@uva.es, ORCID: 
\href{https://orcid.org/0000-0003-0151-8373}{0000-0003-0151-8373}}   
\bigskip
\\
\noindent
$^a$\,Department of Physics, Faculty of Science, Ankara University, 06100 Ankara, T\"urkiye
\\ 
 \noindent
$^b$\,Departamento de F\'{\i}sica Te\'orica, At\'omica y
\'Optica, and IMUVA,\\ Universidad de Valladolid,  47011 Valladolid, Spain
}

\begin{document}

\maketitle

\begin{abstract}
The purpose of this work is to present a method based on the factorizations used in one dimensional quantum mechanics in order to find the symmetries of
quantum and classical superintegrable systems in higher dimensions.
We  apply this procedure to the harmonic oscillator and Kepler-Coulomb systems to show the differences with other more standard approaches.
We have described in detail the basic ingredients to make explicit the parallelism of classical and quantum treatments. One of the most interesting results is the finding of action-angle variables as a natural component of the classical sysmmetries within this formalism.
\end{abstract}



\sect{Introduction}
The aim of this work is to introduce  a method to find the symmetries of quantum and classical superintegrable systems by means of two   paradigmatic examples: the harmonic oscillator (HO) and  Kepler-Coulomb (KC).
The technique we are going to apply is based on the well known factorization method \cite{infeld}, which was already applied in the early times of quantum mechanics \cite{schrodinger40}, although the origin of this type of transformations goes back to Darboux; in that context it is referred to as Darboux transformations and has a wider range of applications in nonlinear equations \cite{salle91}. Another development of factorizations is known as supersymmetric quantum mechanics (SUSY-qm) \cite{cooper,david,oscar}. In principle, the factorization was designed to deal with one dimensional quantum problems; but in the present work we want to show how it can be extended to  higher dimensions and the way that factorizations can  also be used in the classical frame to obtain constants of motion (or ``classical symmetries''; it happens that constants of motion are generating functions of canonical transformations that leave invariant the Hamiltonian). In fact, the symmetries that are obtained in this way for classical systems, can be naturally interpreted in terms of action-angle variables of Hamilton-Jacobi theory, what supplies an  extra interest to  this approach. In general, the methods used to obtain symmetries  for quantum superintegrable systems and their classical analogs follow quite different routes; for instance, in the classical context often it is made use of Hamilton-Jacobi formalism  \cite{post10,kalninskress11,kalnins11} which is not valid for the quantum case. However, as we will show (see also Refs. \cite{negro13,negro14,negro16,negro17}), the factorization method supply a unified way towards the symmetries in classical and quantum  mechanical systems.

Thus, in this work we will pay attention essentially to explain  the ingredients of this method. We have chosen the  HO and KC as two well known  examples \cite{demkov59,fradkin65,fradkin67}, where it is easy to check the consistency of our results and the standard expressions, to show how the method works, what are the key points and at the same time to appreciate the differences. For instance, the symmetries so obtained in the quantum case, are ready to compute eigenfunctions and degeneracy eigenspaces; while in the classical case the constants of motion here found lead directly to  the open and closed trajectories or even to the motion, as well as their frequencies. The  algebraic structure of symmetries in both quantum and classical contexts, can also be found inside this formalism and their similarity is immediate.

The method here presented is designed to deal with superintegrable systems \cite{evans90,winternitz13}. Recall that a system with $n$ degrees of freedom  that has $n$ involutive symmetries in the quantum formalism  (or constants of motion in classical mechanics), including the Hamiltonian if it does not depend explicitly on time, is said to be integrable; when there are additional symmetries (not commuting with all the involutive ones) it is called superintegrable;  The name maximally superintegrable is reserved when  there is  the maximum number $2n-1$  of  independent  symmetries.
In this scenario, the factorization method has been already put to work in  a variety of configurations: systems defined on constant curvature surfaces \cite{negro16,negro17}; to find higher order symmetries \cite{negro13,ttw09,ttw10,perlick17} or to a  list of  classical systems \cite{negro08,hussin19}. In the near future we plan to apply it to further  superintegrable models, such as Smorodinsky-Winternitz \cite{rodriguez01}, Evans \cite{evans08}, or those of Darboux \cite{ballesteros13} and Perlick \cite{perlick17}, in order to show its flexibility for a wide class of problems. Another point of interest regarding this method is that it allows for its generalization to any dimension in a straightforward way.  

The organization of this work is as follows. In Section 2 we introduce the spherical coordinates  appropriate to work with central potentials which is the case of HO and KC. The separation of variables leads to three reduced Hamiltonians in a fixed order, where the first two of them correspond to the angular variables.
Then, the well known operators $\hat{L}^\pm$ corresponding to the $so(3)$ generators which span the angular symmetries are expressed as the product of two reduced operators in the separated variables. These reduced operators are called shift and ladder operators, since one of them changes the eigenvalues of the angular Hamiltonian $\hat{L}_z^2$, while the other changes a parameter of the potential of a second Hamiltonian, $\hat{L}^2$. At the same time, the reduced operators can be identified with factorization operators of  reduced Hamiltonians. The key point is that the symmetries consist in the product of these two types of operators. This structure is repeated in all the other cases. 
In Sections 3 and 4 we construct the symmetries of  KC and HO, respectively, following the same path. Section 5 is devoted  to the classical symmetries of these two systems, and Section 6 to the relation of classical symmetries and action-angle variables. Some conclusions and remarks close this paper.

\sect{Symmetries in Spherical Coordinates of Quantum Central Potentials}

In this section we will introduce the spherical coordinates $(r,\theta,\phi)$, where $r$ is the radius,  $\theta$ is the polar and $\varphi$ the azimuthal angles, to study the symmetries of both harmonic oscillator and Kepler-Coulomb systems under the same approach based on factorizations. Both systems are maximally superintegrable \cite{evans90,winternitz13} and therefore they admit five independent symmetries. 

\subsection{Spherical coordinates}
The Schr\"odinger Hamiltonians for central potentials $V(r)$ have the form of a nested structure in spherical coordinates $(r,\theta,\varphi)$:
\begin{equation}\label{central}
\hat H=-\partial_{rr}-\frac{2}{r}\partial_r +V(r) +\frac{1}{r^2}\left(-\partial_{\theta\theta}-\frac{1}{\tan{\theta}}\partial_\theta+\frac{1}{\sin^2{\theta}}\,\Big(-\partial_{\varphi\varphi}\Big)\right)
\end{equation}
where $2m=\hbar=1$ and $\partial_r = \frac{\partial}{\partial r}$, $\partial_{rr}= \frac{\partial^2}{\partial r^2}$, etc\dots From this structure we can see that the Hamiltonian $\hat H$, the total momentum operator ${\hat L}^2$ and the square of the $z$-component ${\hat L}_z^2$ given by 
\begin{equation}\label{involutive}
{\hat L}_z^2= (-i\partial_\varphi)^2,\qquad
{{\hat L}}^2=-\partial_{\theta\theta}-\frac{1}{\tan{\theta}}\partial_\theta+\frac{{\hat L}_z^2}{\sin^2{\theta}} ,\qquad
\hat H = -\partial_{rr}-\frac{2}{r}\partial_r +V(r) +\frac{{\hat L}^2}{r^2}
\end{equation}
constitute a sequence of three involutive operators, i.e. they commute among each other  and may be interpreted as a kind of partial Hamiltonians. They are associated to the  coordinate separation and determine a set of common eigenfunctions separated into a product of radial and angular components, 
\begin{equation}\label{sepw}
\Psi_{n,\ell,m}(r,\theta,\varphi)=
R_n^{\ell}(r)\Phi_{\ell,m}(\theta,\varphi)=
R_n^{\ell}(r)P^m_{\ell}(\theta) \phi_m(\varphi)
\end{equation}

The physical solutions have additional boundary conditions to be specified later; at this moment we look at the above functions just as separated solutions of  differential eigenvalue equations for this kind of Hamiltonians.
Each component of the solution (\ref{sepw}) satisfies a reduced eigenvalue equation in one variable, for each  of the ``reduced'' partial Hamiltonians (\ref{involutive}) 
consistent with the separation sequence:
\begin{equation}\label{reduced}
\begin{array}{ll}
(a)\quad & {\hat L}_z^2(\varphi) \phi_m(\varphi) := -\partial_{\varphi\varphi}\phi_m(\varphi)= m^2 \phi_m(\varphi)
\\[2.ex]
(b)\quad & {{\hat L}}_m^2(\theta)P_\ell^m(\theta):= \left(-\partial_{\theta\theta}-\frac{1}{\tan{\theta}}\partial_\theta+\frac{m^2}{\sin^2{\theta}}\, \right)
P_\ell^m(\theta) 
= \ell(\ell+1)\, P_\ell^m(\theta)
\\[2.ex]
(c)\quad & {\hat H}_\ell(r)R_n^{\ell}(r): = \left(-\partial_{rr}-\frac{2}{r}\partial_r +V(r) +\frac{\ell(\ell+1)}{r^2}\right) R_n^{\ell}(r) = E_n \, R_n^{\ell}(r)
\end{array}
\end{equation}
where $\ell, m, n$ are quantum numbers to be specified later. For this reason the partial Hamiltonians (\ref{involutive}) are also referred as ``diagonal'' operators in the separated basis (\ref{sepw}).    

Each component eigenfunction depends on the eigenvalue of the  previous reduced Hamiltonian component; for instance in (\ref{reduced}-b), the eigenfunction $P_\ell^m(\theta)$ of $\hat{L}^2$
carries the eigenvalue label $\ell$ of the current reduced operator ${{\hat L}}_m^2$  and the label $m$  of the eigenfunction $\phi_m(\varphi)$, corresponding to the eigenvalue of the previous reduced Hamiltonian ${\hat L}^2_z$. We say that the involutive operators are in a nested construction: ${\hat L}^2_z\subset {{\hat L}}^2\subset {\hat H} $. Therefore, in order to build symmetry operators acting on the separated solutions of the total
Hamiltonian by means of operators associated to the reduced Hamiltonians (\ref{reduced}) they must keep this ``matching'' of eigenvalues of consecutive reduced partial Hamiltonians. 

Our plan is the following. We start with the involutive symmetries
${\hat L}_z^2,\ {\hat L}^2$ and ${\hat H}$, mentioned above, 
then, we want to complete the number of independent symmetries (up to five in three dimensions)  by computing two new pairs: i) $\Lpm{\theta,\varphi}$ depending on the variables $\theta,\  \varphi$, which will commute with the Hamiltonians ${\hat L}^2$ and  ${\hat H}$; and 
ii) $\hat {\mathcal S}^\pm_{r,\theta}$ depending on the variables $r,\ \theta$, which will commute with the Hamiltonians ${\hat H}$ and ${\hat L}_z^2$. We present all of them  in the following natural ordering:
\begin{equation}\label{simetrias}
{\hat H}\, , \quad \Spm{r,\theta}\, , \quad {\hat L}^2\, ,\quad 
\Lpm{\theta,\varphi}\, ,\quad \hat{L}_z^2
\end{equation}
We will arrive to these symmetries through the factorization of the reduced partial Hamiltonians. Of course not all of them will be independent, but we will check that  five of them will be.

\subsection{Angular states and symmetry operators $\Lpm{\theta,\varphi}$}
We will start by analyzing the symmetries of the total angular momentum operator 
${\hat L}^2$
which are well known in quantum mechanics.
Consider the lowering and raising momentum operators ${\hat L}^\pm$ (in our context we will use the notation $\Lpm{\theta,\varphi}$ for them):
\begin{equation}\label{l1a}
\Lpm{\theta,\varphi}= e^{\pm i \varphi}(\pm \partial_{\theta}+i \frac{1}{\tan{\theta}}\partial_\varphi)
\end{equation}
These operators commute with the total angular momentum but they modify the eigenvalues of ${\hat L}_z$,
\begin{equation}\label{l2}
\Lpm{\theta,\varphi}\,  {\hat L}^2 = {\hat L}^2\, \Lpm{\theta,\varphi} \,, \qquad {\hat L}_z\, \Lpm{\theta,\varphi}  = \Lpm{\theta,\varphi}\, ({\hat L}_z\pm1)
\end{equation}
Our goal is to showcase how this structure translate into the separated components of  the angular eigenfunctions shown in (\ref{sepw}):
\begin{equation}\label{phivar}
{{\hat L}}^2\,\Phi_{\ell,m}(\theta,\varphi)=\ell(\ell+1)\,\Phi_{\ell,m}(\theta,\varphi),\qquad
\Phi_{\ell,m}(\theta,\varphi)=P_\ell^m(\theta)\phi_m(\varphi)
\end{equation}

We can split the $\Lpm{\theta,\varphi}$ operators into a product of two one-dimensional factors.

\begin{itemize}

\item[(i)] A factor is one of the pair of exponentials $e^{\pm i\varphi}$ which we rewrite in the form 
\begin{equation}\label{vp}
\lpm{\varphi} := 
e^{\pm i\varphi}
\end{equation}
Then, we say that $\lpm{\varphi}$ are  {\bf ladder operators} of the ``diagonal'' operator ${\hat L}_z$ in the variable $\varphi$. This is clear, since if 
 $m\in \mathbb Z$ designs the eigenvalues of the eigenfunctions $\phi_m(\varphi)= e^{im\varphi}$ of ${\hat L}_z$, then
\begin{equation}\label{aapm}
\begin{array}{c}
{\hat L}_z\phi_{m}(\varphi)= m \phi_{m}(\varphi),\qquad \lpm{\varphi}\phi_{m}(\varphi)\propto \phi_{m\pm1}(\varphi),\qquad
[{\hat L}_z, \lpm{\varphi}] = \pm \lpm{\varphi}
\end{array}
\end{equation}

\item[(ii)] The second pair of factor operators in (\ref{l1a}) are 
\begin{equation}
\dpm{\theta,\varphi}:=\pm \partial_{\theta}+i \frac{\partial_\varphi}{\tan{\theta}}
\end{equation}
If they act on separated eigenfunctions (\ref{phivar}), they will give rise to the  reduced operators $\dpm{\theta,m}$ defined  by:  
\begin{equation}\label{redthetas}
\begin{array}{ll}
\Lm{\theta,\varphi} \,\Big(P_\ell^m(\theta)\phi_m(\varphi)\Big):= \Big(\dm{\theta,m}\,P_\ell^m(\theta)\Big)\phi_{m-1}(\varphi)\,,\qquad 
&\dm{\theta,m}:=- \partial_{\theta}- \frac{ m}{\tan{\theta}}\,
\\[2.5ex]
\Lpp{\theta,\varphi}\,\Big(P_\ell^{m-1}(\theta) \phi_{m-1}(\varphi)\Big):=\Big(\dpp{\theta,m}\,P_\ell^{m-1}(\theta)\Big)\phi_m(\varphi)\,,\qquad 
&\dpp{\theta,m}:= \partial_{\theta}- \frac{ m-1}{\tan{\theta}}
\end{array}
\end{equation}
where $ m$ is for the eigenvalues of $L_z$ as mentioned above.   
The reduced operators $\dpm{\theta,m}$ act on the components $P_\ell^m(\theta)$ of the eigenfunctions $\Phi_\ell^m(\theta,\varphi)$ as follows,
\begin{equation}\label{dtheta}
\dm{\theta,m}\, P_\ell^{m}(\theta) \propto P_\ell^{m-1}(\theta)\,,\qquad
\dpp{\theta,m}\, P_\ell^{m-1}(\theta) \propto P_\ell^{m}(\theta)
\end{equation}

\item[(iii)]
The total angular momentum, written as
\begin{equation}\label{l2m}
{{\hat L}}^2(\theta,\varphi)= -\partial_{\theta\theta}-\frac{1}{\tan{\theta}}\partial_\theta-\frac{\partial_{\varphi\varphi}}{\sin^2{\theta}}
\end{equation}
 when it acts on angular eigenfunctions $\Phi_{\ell,m}(\theta,\varphi)$
  leads to the reduced Hamiltonians ${{\hat L}}_{m}^2(\theta)$ (or ${{\hat L}}_{\theta,m}^2$)  in $\theta$,
\begin{equation}\label{l21}
 {{\hat L}}_{m}^2(\theta) 
 := 
 -\partial_{\theta\theta}-\frac{1}{\tan{\theta}}\partial_\theta+\frac{ m^2}{\sin^2{\theta}}
\end{equation}

The reduced operators $\dpm{\theta,m}$ in the variable $\theta$ are called {\bf pure shift (or displacement) operators} of
the reduced operator ${\hat L}^2_m$ given in (\ref{reduced}-b) since when they act on an eigenfunction component $P_\ell^m(\theta)$, they change
the parameter $m$ according to (\ref{dtheta}),
but the eigenvalue (determined by $\ell$) is left invariant. This is shown by means of the intertwining and factorization relations 
\begin{equation}\label{bbpm}
{\hat L}^2_{m}(\theta) \dpp{\theta, m} = \dpp{\theta,m} {\hat L}^2_{m-1}(\theta),\qquad
\dm{\theta,m}{\hat L}^2_{m}(\theta)  =  {\hat L}^2_{m-1}(\theta) \dm{\theta,m},\qquad
\dm{\theta,m}\dpp{\theta,m} = {\hat L}^2_{m-1}(\theta)- m(m-1)
\end{equation}

The last expression of (\ref{bbpm}) means that, if  the maximum value of $|m|$ is $m_{\rm max}=\ell$,  and the eigenvalues of ${\hat L}^2_{m}$ are $\ell(\ell+1)$ then $-\ell\leq m \leq \ell$ and
\begin{equation}\label{dtheta2}
\dm{\theta,-\ell}\, P_\ell^{-m=-\ell}(\theta) =0\,,\qquad
\dpp{\theta,\ell+1}\, P_{\ell}^{m=\ell}(\theta) =0
\end{equation}

Relations (\ref{bbpm}) for $\dpm{\theta,m}$ are well known as standard factorization of ${\hat L}_m(\theta)$.
Notice that the intertwining (\ref{bbpm}), can also be rewritten in terms of commutation rules,
\begin{equation}\label{bbpm2}
[{\hat L}^2_{m}(\theta), \dpp{\theta,m}] = \dpp{\theta,m}\Big( {\hat L}^2_{m-1}(\theta)-{\hat L}^2_{m}(\theta)\Big),\qquad
[\dm{\theta,m},{\hat L}^2_{m}(\theta)]  =  \Big({\hat L}^2_{m-1}(\theta)-{\hat L}^2_{m}(\theta)\Big) \dm{\theta,m}
\end{equation}

\item[(iv)] In conclusion, we may consider the symmetry operators $\Lpm{\theta,\varphi}$ of (\ref{l1a})  in a reduced form as the product of ladder $\lpm{\varphi}$ (of ${\hat L}_z(\varphi)$) 
and shift $\dpm{\theta,m}$ (of the  following partial Hamiltonian $L_m^2(\theta)$) operators 

\begin{equation}\label{slad}
\Lpm{\theta, m, \varphi}= \dpm{\theta,m} \lpm{\varphi}
\end{equation}
In this case the shift operator can be simplified:
\[
\dpm{\theta,\varphi}=\pm \partial_{\theta}+i \frac{\partial_\varphi}{\tan{\theta}} \Rightarrow \dpm{\theta,m}:=\pm \partial_{\theta}+ \frac{m}{\tan{\theta}}
\]

Its action on separated solutions gives new separated solutions with the same eigenvalue (the same  $\ell$-value):
\[
\Lpp{\theta,\varphi}\, \big(P_\ell^m(\theta)\phi_m(\varphi)\big)\propto
P_\ell^{m+1}(\theta)\phi_{m+1}(\varphi),
\qquad
\Lm{\theta,\varphi}\, \big(P_\ell^m(\theta)\phi_m(\varphi)\big)\propto
P_\ell^{m-1}(\theta)\phi_{m-1}(\varphi)
\]
\end{itemize}
 We have thus shown that the angular momentum symmetries $\Lpm{\theta,\varphi}$ have an structure 
shown in the scheme:

\begin{equation}\label{eschemelpm}
 {\hat L}^2 (\theta)\, \  \to\   \  \Lpm{\theta,   \varphi}=\dpm{\theta }\,\lpm{\varphi}\, \ \leftarrow\  {\hat L}_z^2(\varphi)
\end{equation}

This will be a general  property of the symmetries that we will find in the following:
i) $\Lpm{\theta,  \varphi}$ depends on the variables $\varphi$ and $\theta$ of two consecutive reduced Hamiltonians: $\hat{L}^2_m(\theta)$ and  
${\hat L}^2_z(\varphi)$. ii) The operator $\Lpm{\theta, \varphi}$ is a symmetry of 
${\hat L}^2(\theta,\varphi)$, but its commutator with ${\hat L}_z(\varphi)$ is of ladder type (see (\ref{l2})):
\begin{equation}\label{slad2}
[{\hat L}^2,\Lpm{\theta,  \varphi}]= 0\,,\qquad
[{\hat L}_z,\Lpm{\theta,   \varphi}]= \pm  \Lpm{\theta, \varphi}
\end{equation}

\medskip

\subsection{Ladder operators of ${{\hat L}}^2$}
\medskip

Next, we will use the considerations of the previous section to find the symmetry operators
$\Spm{r,\theta}$, mentioned in (\ref{simetrias}), which will be obtained from a ladder operator
of  the reduced ${\hat L}^2_{m}(\theta)$ and a shift operator of the reduced Hamiltonian ${\hat H}_{\ell}(r)$. Thus, in this case we follow the opposite way: Firstly we  find the  reduced operators by means of the factorization method and from them we construct a kind of symmetry operators.

We start by looking for ladder operators $\lpm{ \ell}(\theta)$ for 
${\hat L}^2_{\theta, m}$ that play the same role as $\lpm{\varphi}$ with respect to ${\hat L}_z$, i.e., modify the eigenvalues of the total angular momentum, $\ell$. Thus, we will consider the reduced eigenvalue equation for ${{\hat L}}_{ m}^2(\theta)$, given by (\ref{l21}), and write it in the following form 
\begin{equation}\label{b}
{\hat C}_{\ell}(\theta)P_{\ell}^{m}(\theta):=\Big(-\sin^2{\theta}\partial_{\theta\theta}-\sin{\theta}\cos{\theta}\partial_\theta-\ell(\ell+1)\sin^2{\theta}\Big)P_{\ell}^{m}(\theta)=-m^2 P_{\ell}^{m}(\theta)
\end{equation}
where equation (\ref{b}) is an eigenvalue equation for a new operator ${\hat C}_{\ell}(\theta)$ with eigenvalues $-m^2$ and where $\ell$ plays the role of a parameter of such operator.
Applying the factorization method to this equation,
we obtain the factor operators \cite{kuru12,hussin}
\begin{equation}\label{lpm}
\lpm{\ell}(\theta)=\pm\sin{\theta}\partial_\theta + \ell\cos{\theta}\,
\end{equation}
which satisfy the following factorization properties
%
\[
\begin{array}{c}
    {\hat C}_\ell(\theta)=\lpp{\ell,\theta} \lm{\ell,\theta}  -\ell^2 =-m^2 
     \\[2.5ex]
     \lm{\ell,\theta} \,{\hat C}_\ell= 
     {\hat C}_{\ell-1}\,\lm{\ell,\theta}\,, \qquad 
\lpp{\ell,\theta}\,{\hat C}_{\ell-1}=
    {\hat C}_{\ell}\,\lpp{\ell,\theta} 
    \end{array}
\]
\[
[\hat L^2_m,\lpm{\ell,\theta}]_{L^2\to\ell(\ell+1)}= \pm 2\ell\,\lpm{\ell,\theta}
\]
This means that $\ell\geq |m|$, and that the action of $\lpm{\ell,\theta}$ on  eigenfunctions $\Phi_{\ell,m}(\theta,\varphi)=P_\ell^m(\theta)\phi_m(\varphi)$ (\ref{phivar}) is to change the eigenvalue parameter $\ell $ of ${\hat L}^2_m(\theta)$ by $\ell\pm1$, while the label $m$ keeps unaltered: 

\[
\lm{\ell}(\theta)\, P^m_{\ell}(\theta)\propto P^m_{\ell-1}(\theta)\,,\qquad
\lpp{\ell+1}(\theta)\, P^m_{\ell}(\theta)\propto P^m_{\ell+1}(\theta)\,,
\qquad \lm{\ell}(\theta)\, P^{m=\ell}_{\ell}(\theta)=0
\]
where the last equation determines the lowest $\ell$-state, $\ell_{\rm min}= m$.  The ladder operators $\lpm{\ell}(\theta)$ of ${\hat L}^2$, together with the symmetry operators $\Lpm{\theta, \varphi}$ generate a spectrum generating algebra (SGA) of ${\hat L}^2$. 

\begin{figure}[h!]
	\centering
\includegraphics[width= 8 cm]{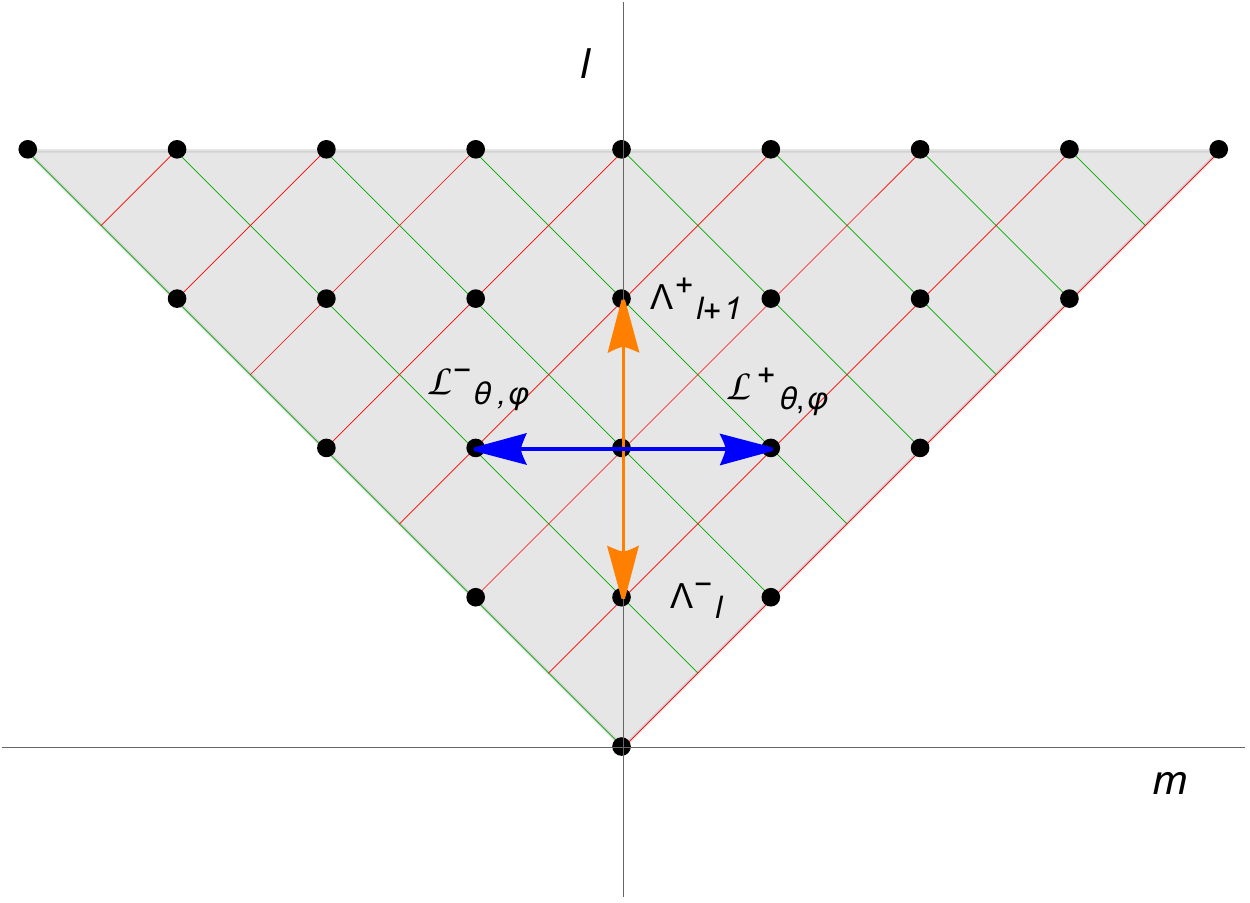}
\caption{Schematic representation of the ladder operators $\lpm{\ell}(\theta)$ and symmetries $\Lpm{\theta, \varphi}$ for the angular Hamiltonian ${\hat L}^2$. Each state is represented by a dot in the plane $(m,\ell)$. \label{fig1}}
\end{figure}

These operators allow us to define the normalizable angular states of the system by acting on a fundamental state  annihilated by ladder and symmetry operators
$\lm{\ell=0}$ and $\Lm{\theta, \varphi}$,
\[
\lm{\ell=0}\Phi_{\ell=0,m=0}=\Lm{\theta,\varphi}\Phi_{\ell=0,m=0}=0\ \implies \Phi_{0,0}\propto 1
\]     
\begin{equation}\label{generating}
\Phi_{\ell, m}(\theta,\varphi)=
N\,(\Lpp{\theta,\varphi})^{m}(\prod_{i=1}^{\ell}\lpp{i,\theta})\Phi_{0,0},\qquad
0\leq m\leq \ell
\end{equation}
where $N$ is a normalization constant. In other words, we can move on different eigenvalue ($\ell$) states by means of ladder operators, and inside each energy level by symmetry operators (changing $m$). This is essentially the action of an spectrum generating algebra \cite{kepler12,negro12} of the angular operator ${\hat L}^2$. The separated angular solutions obtained in this way are spherical harmonics, $\Phi_{\ell, m}(\theta,\varphi)=
Y_\ell^{m}(\theta,\varphi)$.
For instance, 
\[
Y_2^{2}(\theta,\varphi)\propto (\Lpp{\theta,\varphi})^2 \lpp{2}(\theta)\lpp{1}(\theta)  \Phi_{0, 0}(\theta,\varphi).
\]
These angular operators and the angular wavefunctions, are common to all central potential systems.

\sect{Radial Symmetries: The Harmonic Oscillator}

To complete the symmetry analysis, we must consider the reduced radial eigenvalue problem in order to find the remaining shift operator $\dpm{r,\ell}$; then together with the known ladder $\lpm{\ell,\theta}$ operators (see (\ref{lpm})) we will form the second pair of symmetries  following the pattern of the angular Hamiltonian. The radial shift operators will depend on the particular form of the radial potential $V(r)$; however, the ladder operators $\lpm{\ell,\theta}$ of ${\hat L}^2_{\theta, m}$ (which are given above) are common to all central potentials. Thus, we have the following scheme for the remaining symmetry, $\Spm{r,\ell, \theta}$, similar to (\ref{eschemelpm}),
\begin{equation}\label{eschemespm}
{\hat H}_\ell(r)\,\ \to\   \Spm{r,\ell,\theta}=\dpm{r, \ell}\lpm{\ell,\theta}\, \leftarrow \  {\hat L}^2_m(\theta)
\end{equation}

In the following, we will apply the factorization method to the reduced radial Hamiltonian 
\begin{equation}\label{radial}
{\hat H}_{\ell}(r)=-\partial_{rr}-\frac{2}{r}\partial_r+\frac{\ell(\ell+1)}{r^2}+V(r)
\end{equation}
in order to find its shift $\dpm{r,\ell}$ (and ladder $\lpm{r}$) operators.
The central potential $V(r)$ will be
that of the HO or the KC potentials
\begin{equation}\label{radial2}
 V_{\rm HO}=\frac{{\omega}^2}{4}\,r^2, \qquad V_{\rm KC}=\frac{-k}{r}
\end{equation}
where $k$ is a positive real constant of the KC potential;  $\omega>0$ is an angular frequency and $\ell$ determines one of the eigenvalues of ${\hat L}^2$.
 
We will devote this section to finding the shift and ladder operators  
for the HO Hamiltonian ${\hat H}_{\rm HO}$ with potential $V_{\rm HO}(r)$. In the following section we will work out the same method for the
KC potential $V_{\rm KC}(r)$.

\subsection{Shift and ladder operators of harmonic oscilator}

\noindent
{\bf\em Basic operators $\boldsymbol{ {\hat a}^\pm_\ell, {\hat b}^\pm_\ell}$ of
radial oscillator $\boldsymbol{{\hat H}_\ell(r)}$}
\medskip

We will handle the following notation for the reduced HO Hamiltonian, and the corresponding discrete eigenfunctions and eigenvalues: 
\begin{equation}\label{horadial}
{\hat H}_\ell(r)R_n^{\ell}(r)=\left(-\partial_{rr}-\frac{2}{r}\partial_r+\frac{\ell(\ell+1)}{r^2}+\frac{\omega^2}{4}\,r^2\right)R_n^{\ell}(r) = E_nR_n^{\ell}(r)
\end{equation}
Thus, we look for the factorizations of  ${\hat H}_\ell(r)={\hat H}_\ell$. There are two independent factorization sets of radial operators $\{{\hat a}_\ell^\pm, {\hat b}_\ell^\pm\}$ given by \cite{david96}:
\begin{equation}\label{abs}
\begin{array}{ll}
{\hat H}_\ell={\hat a}^+_\ell {\hat a}^-_\ell-\frac{\omega}{2}(2\ell-1), 
& 
\left\{ \begin{array}{ll}
{\hat a}^+_{\ell}=-\partial_r+\frac{\ell-1}{r}+\frac{\omega}{2}r ,
   
    \qquad  &{\hat a}^+_{\ell+1}R_n^\ell\propto R_{n+1}^{\ell+1}  \\ &  \\
   {\hat a}^-_{\ell}=\partial_r+\frac{\ell+1}{r}+\frac{\omega}{2}r, \qquad  & {\hat a}^-_{\ell}R_n^\ell \propto R_{n-1}^ {\ell-1}  \\
\end{array}\right.
 \\[2.ex]  
 \\[2.ex]
{\hat H}_\ell={\hat b}^-_{\ell+1} {\hat b}^+_{\ell+1} +\frac{\omega}{2}(2\ell+3),\quad
&\left\{ \begin{array}{ll}
     {\hat b}^+_{\ell}=-\partial_r+\frac{\ell-1}{r}-\frac{\omega}{2}r, \qquad  & {\hat b}^+_{\ell+1}R_n^\ell \propto R_{n-1}^{\ell+1}  \\[2.ex]
    {\hat b}^-_{\ell}=\partial_r+\frac{\ell+1}{r}-\frac{\omega}{2}r,   & {\hat b}^-_{\ell}R_n^\ell\propto R_{n+1}^{\ell-1}  \\
\end{array}\right.
 \\[1.ex]  
 \end{array}
\end{equation}
Both sets are related by the reflection  $-\ell\rightarrow (\ell+1)$. 
The operators ${\hat a}^\pm_{\ell}$ and ${\hat b}^\pm_{\ell}$ close a kind of two
independent Heisenberg algebras satisfying the following
commutation relations,
\begin{equation}\label{aabbs}
{\hat a}_{\ell+1}^- {\hat a}_{\ell+1}^+ - {\hat a}_{\ell}^+ {\hat a}_{\ell}^- = 2\omega,\qquad
{\hat b}_{\ell+1}^- {\hat b}_{\ell+1}^+ - {\hat b}_{\ell}^+ {\hat b}_{\ell}^- = -2\omega
\end{equation}

\begin{figure}[h!]
	\centering
\includegraphics[width= 7 cm]{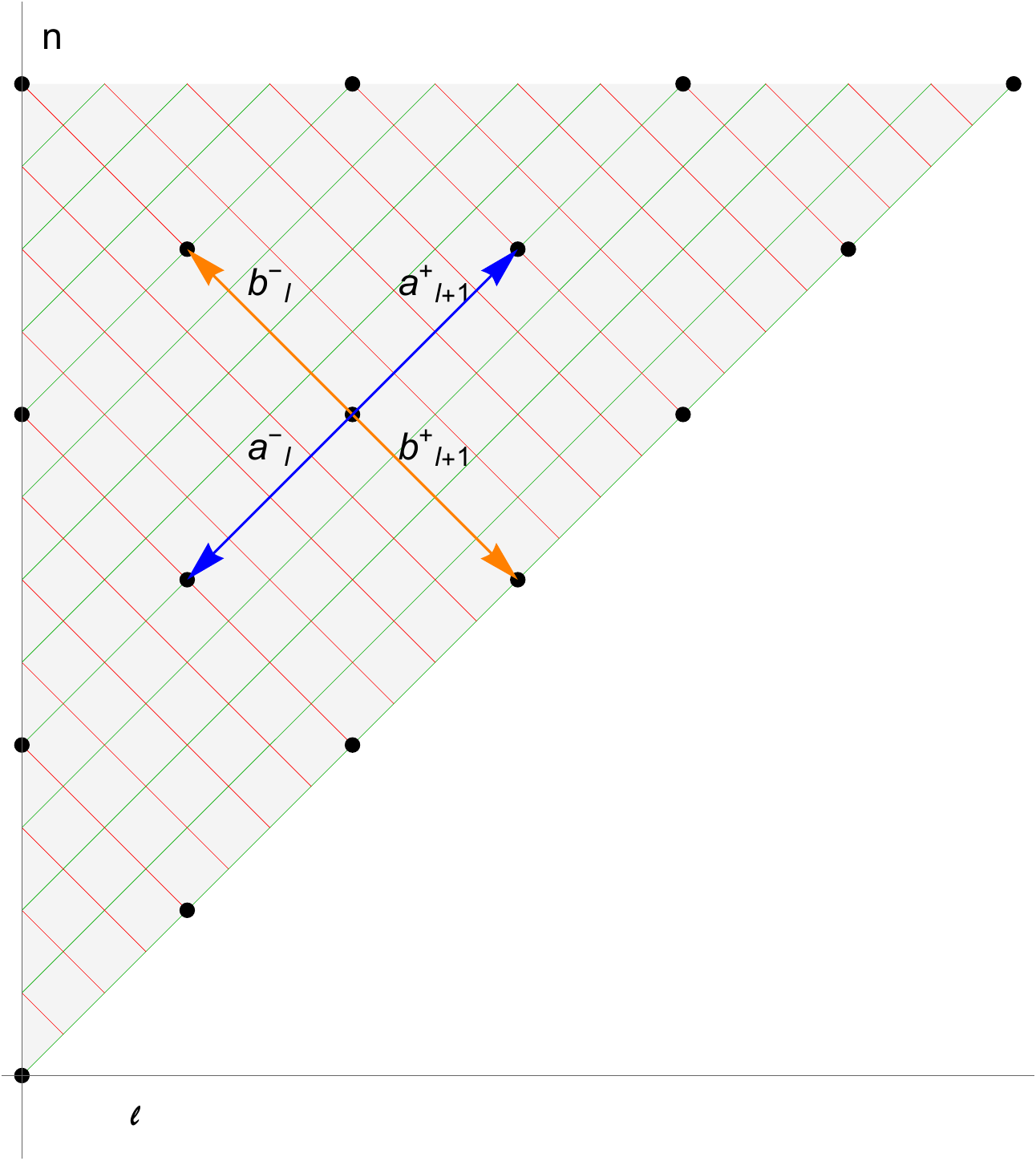}
\caption{Schematic representation of the action of the auxiliary operators ${\hat a}_\ell^\pm(r)$ and ${\hat b}_\ell^\pm(r)$  on functions 
$R_n^\ell(r)$ represented by the points $(\ell, n)$  in the plane. \label{fig2}}
\end{figure}

These elementary factorization operators modify both the energy  ($n$) and the angular momentum  ($\ell$) parameters of the radial states $R^{\ell}_n(r)$, as it is displayed in the commutation rules 
\[
\begin{array}{cc}
  {\hat a}^-_\ell {\hat H}_\ell=({\hat H}_{\ell-1}+\omega){\hat a}^-_\ell, \qquad
   {\hat a}^+_{\ell+1}{\hat H}_\ell=({\hat H}_{\ell+1}-\omega){\hat a}^+_{\ell+1}
   \\[2.ex]
       {\hat b}^-_\ell {\hat H}_\ell=({\hat H}_{\ell-1}-\omega){\hat b}^-_\ell, \qquad
   {\hat b}^+_{\ell+1}{\hat H}_\ell=({\hat H}_{\ell+1}+\omega){\hat b}^+_{\ell+1}\\[1.ex]
\end{array}
\]
where one must be careful because ${\hat a}_{\ell}^-$ is an annihilation operator of energy quantum $\omega$, while that role
is played by ${\hat b}_{\ell}^+$ in the second set.
The effect of these operators is summarized in the following diagram Fig.\ref{fig2} where each point $(\ell,n)$ represents the radial state $R_n^\ell$. 

\noindent
{\bf\em Shift and ladder operators of the radial oscillator $\boldsymbol{{\hat H}_\ell(r)}$}
\medskip

Next, from the previous operators we obtain pure shift, $\dpm{r,\ell}$, that modify only the parameter $\ell$ of angular momentum and pure ladder, $\lpm{\ell,r}$, operators, which change the label $n$ of energy,
\begin{equation}\label{aabb}
\begin{array}{ll}
    \dm{r, \ell}={\hat b}^-_{\ell-1}{\hat a}^-_{\ell}\,,\  & \dpp{r, \ell+2}={\hat b}^+_{\ell+2}{\hat a}^+_{\ell+1}  
  \\[3.5ex]
  \lm{\ell,r }={\hat b}^+_{\ell}{\hat a}^-_{\ell}\,, \   &  \lpp{\ell,r}={\hat b}^-_{\ell+1}{\hat a}^+_{\ell+1}
\end{array}
\end{equation}
They modify their respective parameter in two units, for instance:
\begin{equation}\label{slho}
 \dm{r,\ell} {\hat H}_\ell = {\hat H}_{\ell- 2}\dm{r,\ell}\,,\qquad
\lm{\ell,r} {\hat H}_\ell = ({\hat H}_\ell+ 2\omega) \lm{\ell,r}\,,\qquad
[{\hat H}_\ell(r),\lpm{\ell,r}] = \pm 2 \omega \lpm{\ell,r}
\end{equation}

\begin{figure}[h!]
	\centering
\includegraphics[width= 7 cm]{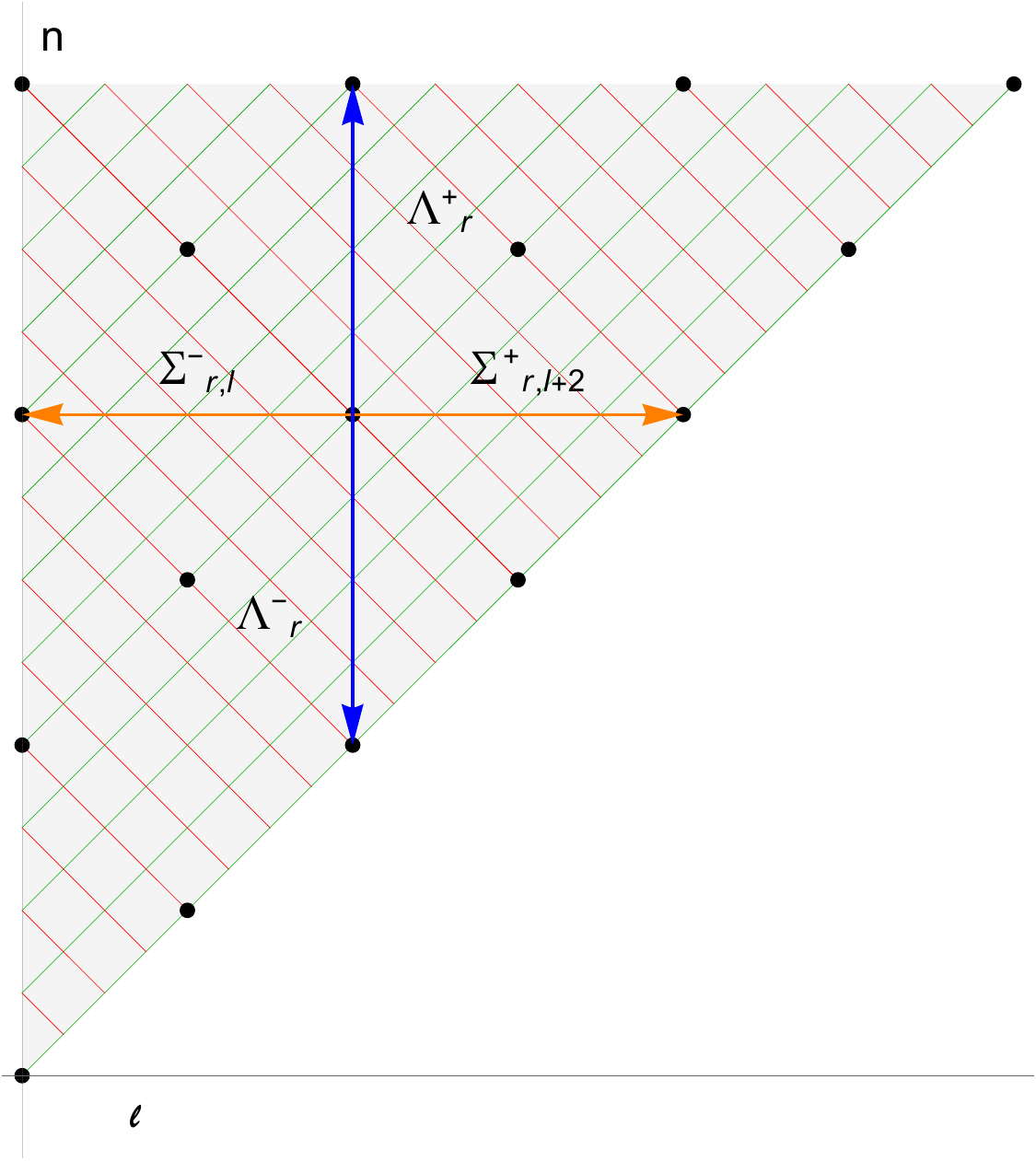}
\caption{Schematic representation of the action of the shift and ladder operators $\dpm{r,\ell}$ and $\lpm{\ell,r}$ on functions 
$R_n^\ell(r)$ represented by the points $(\ell, n)$  in the plane. \label{fig3}
}
\end{figure}

A representation of these operators  can be found in a diagram Fig.\ref{fig3}. For each energy, the state with the maximum value of $\ell$ is annihilated by 
$\hat b^+_{\ell+1}$ (the other options lead to no normalizable eigenstate). The expression ${\hat H}_\ell={\hat b}^-_{\ell+1} {\hat b}^+_{\ell+1} +\frac{\omega}{2}(2\ell+3)$ in (\ref{abs}) leads to the energies: $E_n=\frac{\omega}{2}(2n+3)$, where  $n \geq \ell$. If $\ell=0$, $E_0=\frac{3 \omega}{2} $.

The ladder operators $\lm{\ell, r}={\hat b}^+_{\ell}{\hat a}^-_{\ell}$ and   $\lpp{\ell,r}={\hat b}^-_{\ell+1}{\hat a}^+_{\ell+1}$ depend on $\ell$ through the expression $\ell (\ell+1)$, which can be replaced by $\hat L^2$, therefore they may be rewritten simply as $\lpm{r}$.

Remark that the ladder operators for ${\hat H}_{\ell}$ can not be applied to get 
more symmetries, but they give us the spectrum and the connection of eigenfunctions of different eigenvalues. Therefore, they may be used to build
the spectrum generating algebra \cite{pronko11} of ${\hat H}_{\rm HO}$.

\noindent
{\bf\em Symmetries $\Spm{r,\ell,\theta}$}
\medskip

We have all the ingredients to define the symmetries. For sake of clarity, let us write a complete separated eigenfunction in the form
\[
\Psi_{n,\ell,m}(r,\theta,\varphi) = R_n^{\ell}(r)P_\ell^{m}(\theta)\phi_m(\varphi)
\] 
Thus, in this case the shift operators change two units the value of the parameter $\ell$ of the radial function,
\[
\dpp{r,\ell+2}:R_n^{\ell}(r)P_\ell^{m}(\theta)\phi_m(\varphi) \ \to \ R_n^{\ell+2}(r)P_{\ell}^{m}(\theta)\phi_m(\varphi) 
\]  
Meanwhile, the ladder operators $\lpm{\ell,\theta}$ constructed above acting on the eigenfunctions 
$P_\ell^m(\theta)$,  change the eigenvalue $\ell \to \ell\pm 1$ therefore, we need two of them to match in a new eigenfunction:
\[
\lpp{\ell+2,\theta}\lpp{\ell+1,\theta}:R_n^{\ell}(r)P_\ell^{m}(\theta)\phi_m(\varphi) \ \to \ R_n^{\ell}(r)P_{\ell+2}^{m}(\theta)\phi_m(\varphi) 
\]  
The product of these second order  radial shift and the two angular ladder operators supplies
a second pair of symmetries: 
\[
\Spp{r,\ell+2, \theta} = \dpp{r} \lpp{\ell+2,\theta}\lpp{\ell+1,\theta},\qquad
\Sm{r,\ell, \theta} = \dm{r} \lm{\ell-1,\theta}\lm{\ell,\theta}
\]
The final action of the symmetries $\Spp{r,\ell, \theta}$ is
\[
\Spp{r,\ell+2, \theta}: R_n^{\ell}(r)P_\ell^{m}(\theta)\phi_m(\varphi) \ \to \ R_n^{\ell+ 2}(r)P_{\ell+2}^{m}(\theta)\phi_m(\varphi)\,,\quad
\Sm{r,\ell, \theta}: R_n^{\ell}(r)P_\ell^{m}(\theta)\phi_m(\varphi) \ \to \ R_n^{\ell- 2}(r)P_{\ell-2}^{m}(\theta)\phi_m(\varphi)
\]

They are not independent because their product,
$\Spp{r,\ell, \theta}\Sm{r,\ell, \theta}$ depends on the involutive Hamiltonians. The same happens with $\Lpm{\theta,m,\varphi}$, so at the end we have a total of five independent symmetries, as it should be. 
These symmetries are complemented with the ladder operators
(\ref{slho}) which  connect the eigenspaces with different energy.

Thus, the action of HO symmetries on the separated eigenfunctions is
\begin{equation}
    \Spp{r,\ell,\theta}  \Psi_{n,\ell-2,m}\propto \Psi_{n,\ell,m},\qquad  \Lpp{\theta, m, \varphi}
\Psi_{n,\ell,m-1}\propto \Psi_{n,\ell ,m}
\end{equation}
and the action of ladder operators:
\begin{equation}
\lpm{r} \Psi_{n,\ell,m}\propto \Psi_{n\pm 2,\ell,m} 
\end{equation}
The fundamental state is annihilated by $\lm{r}$ and 
$\Sm{r,\ell,\theta}$. This leads to two solutions (even and odd):
\[
a)\quad \Psi^{\rm e}_0=\Psi_{n=0,\ell=0,m=0}\,,
\qquad
b)\quad \Psi^{\rm o}_0=\Psi_{n=1,\ell=1,m=0}
\]
Then, in this case, there are two sublattices (even and odd) of states $(\ell,n)$ connected by ladder and symmetry operators.
For instance, in the even sector,

\[
\Psi_{2n,0,0}= \left(\lpp{r}\right)^{n} \Psi_0,\qquad
\Psi_{2n,2\ell,0}=\Spp{r,2\ell,\theta}\dots \Spp{r,2,\theta}\Psi_{n,0,0},\qquad
\Psi_{2n,2\ell,m}=(\Lpp{\theta,\varphi})^m\Psi_{2n,2\ell,0}\quad
\]

Next, we will characterize the symmetries of KC in the same spirit.

\sect{Radial Symmetries: Kepler-Coulomb}

In this case, the factorization applied to the reduced Hamiltonian 
${\hat H}_\ell(r)={\hat H}_\ell$ 
\begin{equation}\label{c}
{\hat H}_\ell= 
 - \partial_{rr} -\frac{2}{r}\partial_r+\frac{\ell(\ell+1)}{r^2}-\frac{k}{r}
\end{equation}
lead us to only one pair of shift operators:
\[
\dpm{r, \ell}=\displaystyle
\mp\partial_r+\frac{l\mp 1}{r}-\frac{k}{2\ell} 
\]
That modify  the angular momentum, $\ell$, keeping the energy, $n$,
\begin{equation}
{\hat H}_\ell=\dm{r, \ell+1} \dpp{r, \ell+1}-\frac{k^2}{4(\ell+1)^2},
\qquad
 \left\{ \begin{array}{ll}
    \dpp{r, \ell+1}=-\partial_r+\frac{\ell}{r}-\frac{k}{2(\ell+1)},  & \qquad \dpp{r, \ell+1}R_n^\ell\propto R_{n}^{\ell+1}  \\
     &  \\
    \dm{r, \ell}=\partial_r+\frac{\ell+1}{r}-\frac{k}{2\ell},   
    & \qquad \dm{r, \ell}R_n^\ell \propto R_{n}^{\ell-1}  \\
\end{array}\right.
\end{equation}
Then, for ${\hat H}_\ell$ the energy will be given by 
\[
E_n=-\frac{k^2}{4(n+1)^2},\quad n\geq \ell
\]

\subsection{Coupling: Symmetries of Kepler-Coulomb}

The factorization operators $\dpm{r,\ell}$ modify the value of the total angular momentum from the radial states in one unit, this change is matched by the $\lpm{\ell,\theta}$ operators,
 \[
\Sm{r,\ell,\theta}=\dm{r,\ell}\lm{\ell,\theta}, \qquad \Spp{r, \ell+1, \theta}=\dpp{r, \ell+1}\lpp{ \ell+1,\theta}
\]
These operators allow us to move through states of the Keppler-Coulomb system with the same energy since:
\[ \Spp{r, \ell+1, \theta} \Psi_{n,\ell,m}\propto  \Psi_{n,\ell+ 1,m},
\qquad 
\Sm{r, \ell, \theta} \Psi_{n,\ell,m}\propto \Psi_{n,\ell-1,m}
\]

\subsection{Ladder operators of Kepler-Coulomb}

In order to get the ladder operators of KC system we follow
Ref. \cite{kepler12}. Consider the eigenvalue solution of the reduced KC Schr\"odinger equation
\begin{equation}\label{schc}
{\hat H}_{\ell}\,\psi_{\ell}^n=E_n\,\psi_{\ell}^n,\quad
E_n=-\frac{k^2}{4\,(n+ 1)^2}
\end{equation}
where $n=\ell,\ell+1,\dots$.
To obtain ladder operators for this potential, first we
multiply the Schr\"{o}dinger equation by $r^2$ to become
\begin{equation}\label{sch1c}
{\hat h}_{n}\psi_{\ell}^n=(-r^2\, \partial_r^2 - 2r \partial_r-r^2\,
E_n-k r)\,\psi_{\ell}^n=-\ell(\ell+1)\,\psi_{\ell}^n \,
\end{equation}
We propose the factorization
\begin{equation}\label{hkc}
{\hat h}_{n}=-r^2\, \partial_r^2- 2r \partial_r+
\frac{k^2}{4(n+1)^2}r^2-kr=\lpp{r,n}\lm{r,n} +\omega_{n}
\end{equation}
where
\begin{equation}\label{qap1c}
\textstyle \lpp{r,n} =(-r\,\partial_r+\frac{k}{2(n+1)}r-(n+1 )\,{\hat D}_{n}^{-1},\quad
\textstyle \lm{r,n} ={\hat D}_{n}(r\,\partial_r+\frac{k}{2(n+1)}r-n)
\end{equation}
and
\begin{equation}
\omega_{n}=-n(n+1)
\end{equation}
Now, we want this factorization to fulfill the fundamental  property,
\begin{equation}\label{hkc1}
{\hat h}_{n}=\lpp{r,n}\,\lm{r,n}
+\omega_{n}=\lm{r,n+1}\,\lpp{r,n+1} +\omega_{n+1}
\end{equation}
This is true if we take  ${\hat D}_{n}$  as a dilation operator that satisfy
\begin{equation}\label{d}
\begin{array}{l}
{\hat D}_{n}\,r=\frac{n+1}{n}\,r\,{\hat D}_{n},\qquad
{\hat D}_{n}= \exp[\log\Big((n+1)/n \Big)\, r\partial_r]
\end{array}
\end{equation}
Then, from the relation (\ref{hkc1}) we have the following intertwining relation
\begin{equation}\label{hkc2}
\lm{n,r}\,{\hat h}_{n}={\hat h}_{n-1}\,\lm{n,r}\, , \quad
\lpp{n,r}\,{\hat h}_{n-1}={\hat h}_{n}\,\lpp{n,r}
\end{equation}
which allow us to connect consecutive eigenfunctions by means of  ladder operators,
\begin{equation}
\lm{n,r}:\psi_{\ell}^n \to \psi_{\ell}^{n-1},\quad
\lpp{n,r}:\psi_{\ell}^{n-1} \to \psi_{\ell}^{n}
\end{equation}

Then, the operators $\lpm{\ell,n}$, are just the ladder operators we are looking for with the commutation rules:

\begin{equation}\label{akc}
\displaystyle
\lm{r,n+1}\,\lpp{r,n+1} 
-\lpp{r,n}\,\lm{r,n}=
2(n+1)\,,\qquad
[{\hat H}_{\ell},\hat{\Lambda}^+_{r,n}]=\hat{\Lambda}^+_{r,n}(E_\ell^{n - 1}-E_\ell^n)
\end{equation}

In conclusion, the action of KC symmetries on the separated eigenfunctions
\begin{equation}
    \Spp{r,\ell,\theta}  \Psi_{n,\ell-1,m}\propto \Psi_{n,\ell,m},\qquad  \Lpp{\theta,  \varphi}
\Psi_{n,\ell,m-1}\propto \Psi_{n,\ell ,m}
\end{equation}
together with the ladder operators
\begin{equation}
\lpp{n,r} \Psi_{n-1,\ell,m}\propto \Psi_{n,\ell,m},\qquad
\lm{n,r} \Psi_{n,\ell,m}\propto \Psi_{n-1,\ell,m} 
\end{equation} 
generate the eigentates from the ground state, 
\[
\Spm{r,0,\theta}  \Psi_0 = \Lpm{\theta, 0, \varphi}  \Psi_0= 
\lm{0,r} \Psi_0=0
\]
Then eigenfunction space $\Psi_{n,\ell,m}$ with energy $E_n$ is given by:
\[
\Psi_{n,0,0}=\lpp{n,r}\dots \lpp{1,r}\Psi_0,\qquad
\Psi_{n,\ell,0}=\Spp{r,\ell,\theta}\dots \Spp{r,1,\theta}\Psi_{n,0,0},\qquad
\Psi_{n,\ell,m}=(\Lpp{\theta,\varphi})^m \Psi_{n,\ell,0}\quad
\]

\sect{Symmetries of the Classical HO and KC Systems}

Our next objective is to show that the same method based on factorizations
can be applied to classical central Hamiltonians with an expression, similar to (\ref{central}), but now in terms of canonical coordinates $(r,p_r,\theta,p_\theta,\varphi,p_\varphi)$,
\begin{equation}\label{centralc}
H= p_r^2  +V(r) +\frac{1}{r^2}\left(p_\theta^2+\frac{1}{\sin^2\theta}\,\Big( p_\varphi^2\Big) \right) 
\end{equation}
Its structure is the same as the quantum version (\ref{involutive}) with three functions, 
\begin{equation}\label{involutive2}
L_z= p_\varphi,\qquad
{L}^2= p_{\theta}^2 +\frac{L_z^2}{\sin^2{\theta}}  ,\qquad
H =  p_{r}^2  +V(r) +\frac{L^2}{r^2}  
\end{equation}
where $H$  is the Hamiltonian, $L^2$ the total momentum, and $L_z^2$, the square of the angular $z$-component. These functions  
constitute a sequence of three involutive constants of motion, i.e. their Poisson brackets (PB), between any two of them, including $H$, vanishes. Their constant values  are denoted by $E, \ell^2$ and $m^2$, respectively. These constants of motion  lead to a kind of reduced one dimensional Hamiltonians in  one of the variables $\varphi, \theta, r$ when the rest of variables are replaced by the value of a previous constant of motion:
\begin{equation}\label{reducedc}
\begin{array}{ll}
(a)\quad & L_z^2(\varphi,p_\varphi)   := p_\varphi^2=  m^2
\\[2.ex]
(b)\quad & {L}_m^2(\theta,p_\theta):=  p_\theta^2+\frac{m^2}{\sin^2{\theta}}\,  
= \ell^2 
\\[2.ex]
(c)\quad & H_\ell(r,p_{r}): =  p_{r}^2 +V(r) +\frac{\ell^2}{r^2}    = E   
\end{array}
\end{equation}
We start with these involutive constants of motion
$L_z^2,\ L^2$ and $H$; 
then, we will  construct additional constants of motion to reach the maximum number of five, corresponding to maximal superintegrable systems \cite{winternitz13} as it is the case of HO and KC.  We will search these extra constants in the form of two complex pairs: i) the first complex pair, $\Lpmc{\theta,\varphi}$, will depend on the variables $\theta,p_\theta;  \varphi,p_\varphi$. Its PB with  $L^2$ will vanish and consequently its PB with $H$ will also vanish (since $H$ depends on angular variables through $L^2$); and ii) the second pair 
 ${\mathcal S}^\pm_{r,\theta}$ depending on $r, p_r; \theta, p_\theta$, which will commute with  $H$ (and, trivially, with $L_z^2$, because it is independent of $\varphi$ and $p_\varphi$); thus, the same ordering scheme (\ref{simetrias}) is valid here:
\begin{equation}{\label{fimp}}
 H\, , \quad \Spmc{r,\theta}\, , \quad L^2\, ,\quad 
\Lpmc{\theta, \varphi}\, ,\quad L_z^2
\end{equation}
Of course, not all these constants will be independent; we will see how to choose five of them. We will arrive to these symmetries through the factorization of the reduced Hamiltonians (\ref{reducedc}), in terms of of ladder and shift functions.
The question we have to clarify is the definition of ``ladder and shift functions" in the classical frame. We will introduce them along the  examples of HO and KC systems.

\subsection{Angular symmetry functions $ \Lpmc{\theta ,\varphi}$}
 Consider the classical version to the angular symmetries (\ref{l1}) of $L^2$,
\begin{equation}\label{angularc}
    \Lpmc{\theta,\varphi}= e^{\pm i \varphi}(\pm i p_{\theta}-\frac{L_z}{\tan{\theta}})
\end{equation} 
Similarly to the quantum case, these functions  split into   $\varphi$ and   $\theta$ components.
\begin{itemize}

\item[(i)] The $\varphi$-factor: {\bf angular ladder functions} $\lpmc{\varphi}$

The first component corresponds to a pair of ladder functions for the constant of motion $ L_z$:
\begin{equation}\label{aapmc}
\lpmc{\varphi} := 
e^{\pm i\varphi},\quad \{L_z,\lpmc{\varphi}\} = \mp i\, \lpmc{\varphi}
\end{equation}
In the classical frame we say that  a pair of functions 
(like $\lpmc{\varphi}$) is a set of ladder functions for a  constant of
motion (like $L_z$) if a PB like (\ref{aapmc}) is satisfied (which is the classical analog of the commutator (\ref{aapm})).

\item[(ii)] The $\theta$-factor: {\bf angular shift functions} $\spmc{\theta,m}$

Let ${L}_{m}^2(\theta)$ be  the reduced Hamiltonian given in (\ref{reducedc}), then we define the following functions 
\begin{equation} \label{l1}
\spmc{\theta,m}:=\pm i p_{\theta}- \frac{ m}{\tan{\theta}}
\end{equation}
These functions are called shift-functions of ${L}_{m}^2$ since they are characterized by  a factorization relation,
\begin{equation}\label{bbpm2c}
\spc{\theta,m}\smc{\theta,m} = L^2_{m}(\theta)- m^2
\end{equation}
which means that the constant of motion $ L^2_{m}(\theta)$ must have a value $\ell^2$ greater or equal than the constant $m^2$.
The following type of PB, is similar to the commutation (\ref{bbpm2})
\begin{equation}\label{bbpm2c}
\{L^2,\spmc{\theta,m}\}_{p_\varphi\to m}  =  \pm i\frac{2m}{\sin^2\theta}\, \spmc{\theta,m}=
\pm i\frac{\partial L_m^2}{\partial\,m}\, \spmc{\theta,m}
\end{equation}
where the condition $p_\varphi\to m$ in the PB means that after taking the PB the replacement of the constant of motion $p_\varphi$ by its value $m$ sould be made.

The PB with $L^2$ will also be useful later,
\begin{equation}\label{aapmc2}
\{L^2\,,\lpmc{\varphi}\}_{p_\varphi\to m} 
=
\mp i\,\frac{2m}{\sin^2\theta} \lpmc{\varphi}
\end{equation}

\item[(iii)] Then, the {\bf angular constant of motion} $\Lpmc{\theta \varphi}$ of (\ref{angularc})
 can be expressed as 
 \begin{equation}\label{angularsym}
 \Lpmc{\theta,m,\varphi}= \spmc{\theta,m} \lpmc{\varphi}
 \end{equation}
 when the classical system is characterized by $L_z=m$. In other words, it is the product of a ladder function of $L_z^2$ and a shift function of $L_m^2$. 
From (\ref{aapmc2}) and (\ref{bbpm2c}) we check that $\Lpmc{\theta,m,\varphi}$ will commute with $L^2$ and $H$, but they will act as ladder functions of $L_z$, due to (\ref{aapmc}),
\[
\{L^2,\Lpmc{\theta,m,\varphi}\}_{p_\varphi\to m}=\{H,\Lpmc{\theta,m,\varphi}\}_{p_\phi\to m}=0
\,,\qquad \ 
\{L_z,\Lpmc{\theta,m,\varphi}\}_{p_\varphi\to m}=\mp i\Lpmc{\theta,m,\varphi}
\]

We can say that the complex functions $\Lpmc{\theta,m,\varphi}$ constitute a pair of complex symmetries
for $L^2$ (and for $H$) when the value of the constant of motion $L_z$ is $m$. These symmetries are complex conjugate,
$\Lmc{\theta,m,\varphi}=(\Lpc{\theta,m,\varphi})^*$.

Let us remark that in this case, the constant of motion $m$ is the value of $p_\varphi$, which is a polynomial function, therefore $m$ can be replaced by
$p_\varphi$ simplifying the notation and PBs:
\[
\spmc{\theta,m} \to \spmc{\theta,\varphi}:=\pm i p_{\theta}- \frac{ p_\varphi}{\tan{\theta}},\qquad
\Lpmc{\theta,m,\varphi} \to \Lpmc{\theta,\varphi}:= \spmc{\theta,\varphi} \lpmc{\varphi}
 \]

\item[(iv)]
The {\bf ladder functions of the total angular momentum} $L^2$ are obtained by another factorization of $L_m^2$ (following the way shown in Sect-2.3) in the form, 
\begin{equation}\label{ladang}
\lpmc{\ell,\theta}=\pm i \sin{\theta}p_\theta + \ell\cos{\theta}\,,      \quad \lpc{\ell, \theta} \lmc{\ell,\theta} =L_m^2-m^2 = \ell^2-m^2
\end{equation}
where $\ell^2$ is the  value of the constant of motion ${L^2}$. 
An important remark is that similarly to the quantum case the ladder character can be better expressed in terms of the square root $\sqrt{L^2}$:
\begin{equation}\label{ltheta}
  \{ {L^2} , \lpmc{\ell,\theta}\}_{L^2\to \ell^2}= \mp i 2\ell
  \lpmc{\ell,\theta}\,,
  \qquad
  \{\sqrt{L^2} , \lpmc{\ell,\theta}\}_{L^2\to \ell^2}= \mp i
  \lpmc{\ell,\theta}
\end{equation}

\end{itemize}

\subsection{Radial symmetries and ladder functions of the HO}

Next, we will obtain the radial shift  functions associated to the HO system $H_\ell$ in order to find the second pair of complex symmetries
$\Spmc{r, \theta}$ in the variables $r$ and $\theta$. It will also be  useful to find the ladder functions of $H_\ell$, since they will allow us to find the motion, the frequency and a pair of action-angle variables. 

\begin{itemize}

\item[(i)] {\bf Basic factorization functions $a_\ell^\pm, b_\ell^\pm$}. 
We will deal with the reduced HO Hamiltonian (where we have set $2m=1$) 
\begin{equation}\label{horadialc}
H_\ell(r)= p_{r}^2 +\frac{\ell^2}{r^2}+\frac{\omega^2}{4}\,r^2  
\end{equation}
Then we look for the factorizations of this reduced Hamiltonian. There are two independent factorization sets of radial functions $\{a_\ell^\pm, b_\ell^\pm\}$\,, 
\begin{equation}
\begin{array}{ll}
H_{r,\ell}=a_\ell^+ a_\ell^--{\omega}\, \ell\,,
& 
a_\ell^\pm=\mp i p_r+\frac{\ell}{r}+\frac{\omega}{2}\,r 
\\[2.ex]
H_{r,\ell}=b^-_{\ell}   b_{\ell}^+ + {\omega}\,\ell \,,\quad
 &
b_\ell^\pm=\mp i p_r+\frac{\ell }{r}-\frac{\omega}{2}\,r 
\end{array}
\end{equation}

\noindent
\item[(ii)] {\bf Shift and ladder functions}. 
Next, from the previous auxiliary functions $\{a_\ell^\pm, b_\ell^\pm\}$ found in (i) we obtain {\bf pure shift} functions, $\spmc{r,\ell}$, 
in a similar way to (\ref{aabb})
\begin{equation}\label{ladshiftc}
    \spmc{r,\ell }=b^\pm_{\ell} a_{\ell}^\pm 
\end{equation}
The shift functions have the PBs of shift style (similar to the quantum version (\ref{slho})): 
 \[
 \{ H, \spmc{r,\ell}\}_{L^2\to \ell^2} = \pm i \frac{4\ell}{r^2}\, \spmc{r,\ell}=\pm 2 i \frac{\partial H_\ell }{\partial \ell}\, \spmc{r,\ell}\]
They also allows  to get {\bf pure ladder} $\lpm{\ell, r}$ radial functions, 
 in the following way
\begin{equation}\label{ladshiftc2}
      \lpmc{\ell,r }= a^\pm_\ell b^\mp_\ell\,, 
\end{equation}
The radial ladder functions have the following ladder-like PB:
\begin{equation}\label{laddos}
\{H,\lpmc{\ell,r}\}_{L^2\to \ell^2}  = \mp 2 i \omega \, \lpmc{\ell,r}
\end{equation}

This kind of ladder functions $\lpmc{\ell,r }$ depend only on $\ell^2$ which coincides with the eigenvalues of $L^2$. Thus, we can make this replacement in $\lpmc{\ell,r }$; in this case we use the notation $\lpmc{r }$. Although $\lpmc{r }$ are not symmetries,  their PBs (\ref{laddos}) are quite important. This will provide the frequency of the system, and one pair of appropriate canonical variables, as we will see later.

\item[(iii)] {\bf Symmetry functions}.
The product of the second order radial shift, $\spmc{r}$, and angular ladder, $(\lpmc{\theta})^2$, functions, defined in (\ref{ladang}),   supply
a second pair of radial  symmetries:
\begin{equation}\label{csym}
\Spmc{r,\ell, \theta} = \spmc{r,\ell} (\lpmc{\ell,\theta})^2 =
b_\ell^\pm a_\ell^\pm
\Big(\pm i \sin{\theta}p_\theta + \ell \cos{\theta}\Big)^2
\end{equation}
\[
\{H,\Spmc{r,\ell, \theta}\}_{L^2\to \ell^2}  = 0
\]
Due to (\ref{ltheta}), these symmetries have the following PB with
$\sqrt{L^2}$,
\begin{equation}\label{s2}
\{\sqrt{L^2},\Spmc{r,\ell, \theta}\}_{L^2\to \ell^2}  = \mp 2i  \Spmc{r,\ell, \theta}
\end{equation}

\item[(iv)] {\bf Polynomial symmetries}
The symmetries that we are calculating, for instance $\Spmc{r,\ell, \theta}$, have two drawbacks: a) They are not polynomial in the momenta (due to the square root $\ell = \sqrt{L^2}$); and b) they are complex. These two difficulties are solved by taking the linear symmetric and antisymmetric (and dividing by $i$) combination  of the pair of symmetries (or, essentially,  taking the real and the imaginary part of any of them): 

\begin{equation}\label{sim}
    \sipc{r,\theta} =\frac 12 \left(
\Spc{r,\ell,\theta}+\Smc{r,\ell,\theta} \right), \qquad \anpc{r,\theta}=\frac{1}{2i\ell} \left(
\Spc{r,\ell,\theta}-\Smc{r,\ell,\theta} \right)
\end{equation}
They the following form:
\[
\anpc{r,\theta}=(8 m^2 p_r r + 8 \ell^2 p_r 
    p_\theta (-4 \ell^2 + 4 p^2 r^2 + \omega^2 r^4) \sin 2 \theta))/(4 r^2)
\]
\[
\sipc{r,\theta}= ((2 \ell^2 - E r^2) (m^2 + \ell^2 \cos 2 \theta) + 
  2 \ell^2 p_r p_\theta r \sin 2 \theta)/r^2
\]

They are fourth $(\sipc{r,\theta})$ and third $(\anpc{r,\theta})$ degree polynomials in the momenta $p_r, p_\theta$. They can also be expressed in terms of the  $Q_{ij}$ components of the Fradkin-Demkov tensor \cite{demkov59,fradkin65}:
\[
Q_{ij}=p_i p_j+\frac{\omega^2}{4} x_i x_j
\]
Leading to the expressions:
 \begin{equation}{\label{simho}}
   \begin{array}{cc}
     \sipc{r,\theta}=(L^2-L_z^2)H-2 L^2Q_{zz}     &  \\[2ex]
     \anpc{r,\theta}= 2(L_x Q_{yz}- L_yQ_{zx}) & 
    \end{array}
\end{equation}

\end{itemize}

\subsection{Radial symmetries of the classical KC system}

Next, we will write the classical functions relevant to form the
specific symmetries $\Spmc{r,\ell, \theta}$ of the KC system.

\begin{itemize}

\item[(i)] {\bf Shift functions}. In this case, {\bf the factorization} applied to the reduced Hamiltonian 
$H_\ell(r)$ 
\begin{equation}\label{ckc}
H_\ell=  
 p_{r}^2 +\frac{\ell^2}{r^2}-\frac{k}{r}
 \end{equation}
lead us to one pair of shift functions:
\begin{equation}\label{sr}
\spmc{r,\ell}=
\mp\, i\, p_r+\frac{\ell}{r}-\frac{k}{2\ell}\,, \, \qquad \spc{r,\ell}\,\smc{r,\ell }= H+\frac{k^2}{4 \ell^2}
\end{equation}
They have shift-like PBs:
\begin{equation}\label{ssr}
\{H,\spmc{r,\ell }\}= \pm i
\,  \frac{2\ell}{r^2}\spmc{r,\ell }=\pm i
\,  \frac{\partial H_\ell}{\partial \ell}\,\spmc{r }
\end{equation}

\item[(ii)] {\bf Symmetries}.
Similarly to the quantum case,
 \[
\Spmc{r,\ell,\theta}=\spmc{r,\ell}\lpmc{\ell,\theta},\qquad 
\Spmc{r,\ell,\theta}=(\mp i  p_r+\frac{\ell}{r}-\frac{k}{2\ell})\Big(\pm i \sin{\theta}p_\theta + \ell\cos{\theta}\Big)
\]

\item[(iii)] {\bf Ladder functions}, $\lpmc{r }$, of the classical KC system. In the same
way as we did for the quantum ladder operators, we  obtain the ladder functions,
\begin{equation}\label{qap1c}
\textstyle \lpmc{r } =(\mp r\,i p_r+r\,\sqrt{-H}-\frac{k}{2\,\sqrt{-H}} )\,D_{H}^{\mp},\quad
\end{equation}
where
\[
D_{H}^{\mp}= \exp[\mp 2i r p_r \sqrt{-H}/k]
\]
The functions $\lpmc{r }$, are just the ladder functions we are looking for:
%
\begin{equation}\label{akc}
\displaystyle
\{\lpc{r},\lmc{r }\}= \frac{i \,k}{ \sqrt{-H } }\,,\qquad\quad
\{H,\lpmc{r }\}=\mp i\,4 H \sqrt{-H}\,\lpmc{r }\equiv
\mp i\,\alpha(H)\lpmc{r },\quad \  
\end{equation}
where
\[
\alpha(H)= 4 (-H)\sqrt{-H}/k
\]

In the same way as we mentioned in the ladder function of the HO, this
function supply us with the frequency of the motion, $\alpha(H)$ as we will see later. Such frequency depends on the energy, as opposed to the HO, where it is constant.

\item[(iv)] {\bf Polynomial symmetries}
In principle the symmetries $\Spmc{r,\ell,\theta}$ are not polynomial in the momenta $p_r, p_\varphi$, due to the presence of $\ell= \sqrt{L^2}$, and they are complex.
We can obtain real polynomial symmetries in the momentum functions
by means of the symmetric and antisymmetric combination of $\Spc{r,\theta}$
and $\Smc{r,\theta}$:
\begin{equation}{\label{sim}}
    \sipc{r,\theta} =\frac 12 \left(
\Spc{r,\theta}+\Smc{r,\theta} \right), \qquad 
\anpc{r,\theta}=\frac{\ell}{2i} \left(
\Spc{r,\theta}-\Smc{r,\theta} \right)
\end{equation}
\[
\sipc{r,\theta} = \cos (\theta ) \left(\frac{\ell^2}{r}-\frac{k}{2}\right)+{p_\theta } {p_r} \sin
   (\theta )  \,,\qquad
\anpc{r,\theta} = \sin (\theta ) \left(\frac{\ell {p_\theta}}{r}-\frac{k {p_\theta }}{2 \ell}\right)-\ell
   {p_r} \cos (\theta )
\]

They are polynomials in the momenta $p_r, p_\varphi$ of second  and third  degrees , respectively. Furthermore they are related to the well known Runge-Lenz ${\bf A}$ and angular momentum ${\bf L}$ vectors:  
\begin{equation}{\label{simkc}}
\sipc{r,\theta}=A_z,\qquad \anpc{r,\theta}=(L_x A_y-L_y A_x)= {\bf L} \times {\bf A}|_z
\end{equation}
Where the Runge-Lenz vector is:
\[
{\bf A}={\bf p}\times {\bf L}- \frac{k}2\,{\bf r}
\]

\end{itemize}

\sect{Action-Angle Variables and Symmetries}

We collect in Table \ref{tabla} some of the basic factorization functions (ladder, shift, or symmetries) and split them into their modulus and phase. We will make use of them in order to compute 
canonical variables that  correspond to the solutions of the characteristic
function of Hamilton-Jacobi and to action-angle variables.
In this way, we want to stress that such canonical variables take
a natural part in the symmetries (obtained by this method) of superintegrable systems, as we will show below by means of
 HO and KC systems.

\medskip

\begin{table}[h]
\begin{tabular}{|c|c|c|c|c|c|}
\hline

Function & $\Lpmc{\theta,m,\varphi}$ & $\lpmc{\ell,\theta}$ & $a_\ell^\pm$\ (HO)\  & $b_\ell^\pm$ \ (HO)\  & $\spmc{r,\ell } $\ (KC)\ 
\\[2.ex] \hline
Expression
 & $e^{\pm i\,\varphi}(\mp i\,p_\theta-\frac{m}{\tan{\theta}})$& 
$\pm i p_\theta+{\ell}\cos{\theta} $ & $\mp i p_r+\frac{{\ell}}{r}+\frac 12 \omega r $ & $ \mp ip_r+  {\ell}/r-\frac 12 \omega r $ & $\mp i p_r r+\frac{\ell}{r}-\frac{k}{2\ell}$
\\[2ex] \hline
$|\cdot|^2$ &   $\ell^2-m^2$ &   $\ell^2-m^2$ &  $E+\omega \ell$ &   $E-\omega \ell$ & $E+\frac{k^2}{4\ell^2} $
 \\[2ex] \hline
   Phase & $\varphi-\arccos \frac{m \cot{\theta}}{\sqrt{\ell^2-m^2}}$ & 
   $\arccos{\frac{\ell}{\cos{\theta}}}\sqrt{\ell^2-m^2}$ & 
  $ \arccos{\frac{2\ell+\omega r^2 }{2r \sqrt{E +\omega\ell}}}$ &  
$ \arccos{\frac{2\ell-\omega r^2 }{2r \sqrt{E -\omega\ell}}}$ & $ \arccos{\frac{\ell-k r }{r \sqrt{4 \ell^2 E +k^2}}}$ 
  \\[2ex] \hline
\end{tabular}
\caption{ \label{tabla} Some functions together with their modulus and phase, where ${ L}^2 = \ell^2$, $L_z^2 = m^2$.} 
\end{table}

In this section, we want to interpret the constants of motion that we have previously obtained for the classical KC and HO systems. We have seen that, essentially, the complex constants of motion
${\cal S}^\pm_k$, which are related to each pair, $H_k$ and $H_{k+1}$,  of reduced Hamiltonians (in our case, $H_1\equiv H_\varphi$, $H_2\equiv H_\theta$, $H_3 \equiv H_r$), have the structure
\[
{\cal S}^\pm_k= \Sigma_{k+1}^\pm \, \Lambda^\pm_k
\]
where $\Sigma^\pm_{k+1}$ are shift functions of $H_{k+1}$, while $\Lambda^\pm_k$ are ladder functions of $H_k$.
In polar coordinates we write
\[
{\cal S}^\pm_k= |{\cal S}^\pm_k| e^{\pm i\,\xi_{k}}=
|\Sigma^\pm_{k+1}| e^{\pm i\,\sigma_{k+1}}\,|\Lambda^\pm_k| e^{^\pm\,i \lambda_{k}}
\]
The modulus of the symmetry $|{\cal S}^\pm_k|=|\Sigma^\pm_{k+1}||\Lambda^\pm_k|$ depends on the involutive constants of motion, i.e., of the partial Hamiltonians, so it is not an independent symmetry; but the phase $\xi_{k}$ is a  symmetry which is independent of the reduced Hamiltonians. If we consider the pair
\[
( H_k\,,\xi_k = \sigma_{k+1}+\lambda_k) 
\]
It happens that ${\cal S}^\pm_k$ is a symmetry of $H_{k+1}$ but its PB
with $H_{k}$ has the form (see for instance (\ref{s2}),
\begin{equation}\label{angle1}
\{ H_k, {\cal S}^\pm_k \} = \mp i a {\cal S}^\pm_k
\end{equation}
(where $a=1$ for KC and $a=2$ for HO) and, taking into account that ${\cal S}^\pm_k= |{\cal S}^\pm_k| e^{\pm i\,\xi_{k}}$ with 
$\{H_k, |{\cal S}^\pm_k|\}=0$, then
\begin{equation}\label{angle2}
\{ \xi_k, H_k  \} = a 
\end{equation}

Thus, based on these considerations, we expect that the pairs
\begin{equation}\label{angle3}
( H_\varphi,\xi_\varphi),\ ( H_\theta,\xi_\theta)
\end{equation}
are a good choice as starting point to search for  two pairs of canonical variables which are made of constants of motion. They can be modified slightly to get of a set of action-angle canonical coordinates. In fact the $\xi_k$ variables will be identified with the angle, while partial Hamiltonian $H_k$ will be a function of the action variable $J_k$.
The remaining pair $(J_r,\xi_r)$ comes from  the reduced
Hamiltonian $H_r$ and its ladder functions $\Lambda_r^\pm$. Next, we
will  see the details for KC and HO systems.
\medskip

\subsection{Action-angle variables for the KC system}

\begin{itemize}

\item[(1)]
From the pair $(L_z\equiv p_\varphi, \xi_\varphi)$ to the canonical variables $(J_\varphi\equiv p_\varphi, \xi_\varphi)$.  

The constant of motion $\xi_\varphi$ according to
(\ref{angularsym}) and (\ref{aapmc})-(\ref{l1}) is defined by
\begin{equation}
\xi_\varphi= \lambda_\varphi + \sigma_\theta   = \varphi + \arccos\Big(\frac{L_z}{\sqrt{L^2 -L_z^2}} \cot \theta\Big)= \varphi + \arccos(\cot i \cot \theta)
\end{equation}
where the angle $i$ is that of the plane of the motion and the $z$ axis.
The  canonical bracket $\{\xi_\varphi, p_\varphi \}=1$ is fulfilled. Then, in this case, trivially, $p_\varphi$ and $J_\varphi$ coincide, and the $\varphi$ canonical variables are
\begin{equation}
(J_\varphi:= p_\varphi, \xi_\varphi)
\end{equation}

\item[(2)]
From the pair $(\sqrt{L^2}, \xi_\theta)$ to canonical variables $(J_\theta\equiv \sqrt{L^2}, \xi_\theta)$. 

According to (\ref{sr}) and
(\ref{ssr}), we have
\begin{equation}
\xi_\theta =  \lambda_\theta + \sigma_r  = 
\arccos(\csc i \,\cos \theta) +
\arccos \frac{2L^2/r -k}{\sqrt{4L^2 H + k^2}}\,,\qquad \ 
\frac{\sqrt{L^2}}{\sqrt{L^2 -L_z^2}} = \csc i
\end{equation}
From (\ref{ltheta}), we have the  canonical bracket  $\{\xi_\theta, \sqrt{L^2} \}=1$. In conclusion, $J_\theta(L^2) = \ell- J_0$, such that for the minimum value of $L^2$, $J_\theta(m^2) = 0$, then the
action is $J_\theta(L^2) = \ell- m$, where $m$ is the action of the variable $\varphi$.
Finally, 
the second pair of canonical variables is
\[
(J_\theta= \ell - m,\, \xi_\theta)
\]

The PBs of
cross variables of both set of pairs will vanish.

\item[(3)]
The pair $(H_{\rm KC}, \xi_r)$ and the canonical variables $(J_r(H_{\rm KC}), \xi_r)$.  

In this case, we dispose only of a pair of ladder functions to build the last pair of canonical variables $(J_r(H_{\rm KC}), \xi_r)$.
According to (\ref{qap1c}) they are given by
\[
\textstyle \lpmc{r} =(\mp i\, r\,p_r+r\,\sqrt{-H_{\rm KC}}-\frac{k}{2\,\sqrt{-H_{\rm KC}}} )\,\exp[\mp 2i r p_r \sqrt{-H_{\rm KC}}/k]
\]
therefore
\[
\lpmc{r} = |\lpmc{r}| \exp[\pm i \xi_r],\quad
|\lpmc{r}|=\sqrt{\frac{k^2}{-4H_{\rm KC}} - L^2},
\quad \xi_r = 2 rp_r \sqrt{-H_{\rm KC}}/k + \arccos\frac{-k-H_{\rm KC} r}{\sqrt{k^2-4H_{\rm KC} L^2}} 
\]
Recall  the commutator
\[
\{H_{\rm KC},\lpmc{r}\}=\mp i\,4 (-H_{\rm KC}) \sqrt{-H_{\rm KC}}\,\lpmc{r}= \mp i\alpha(H_{\rm KC})\,\lpmc{r}\ 
\]
then, since $\{H_{\rm KC},e^{a \xi}\}=a e^{a \xi} \, \{H_{\rm KC}, \xi\}$,  we get
\begin{equation}\label{timeCK}
\{\xi_{r},H_{\rm KC}\} = \alpha(H_{\rm KC}) =  -4 H_{\rm KC} \sqrt{-H_{\rm KC}}/k
\end{equation}
This PB leads to the canonical variable $J_r$:
\[
\{\xi_r, J_r(H_{\rm KC})\} = 1 \ \implies \ \frac{d\,J_r(H_{\rm KC})}{d\,H_{\rm KC}} \,(-4 H_{\rm KC} \sqrt{-H_{\rm KC}}/k)=1\ \implies\ 
J_r = \frac{k}{2\sqrt{-H_{\rm KC}}} - \ell\,, 
\]
where $\ell$ is the constant of integration associated to  the value
$J = 0$ at the minimum value of the potential well.
\end{itemize}

In summary, the symmetries allow us to find the canonical variables
\begin{equation}\label{aakc}
(J_\varphi=p_\varphi, \xi_\varphi;\, J_\theta= \ell- m, \xi_\theta;\,  J_r= \frac1{2\sqrt{-E}} - \ell,\xi_r)
\end{equation}
such that all of them are constants of motion except
$\xi_r$, whose evolution is given by (\ref{timeCK}),
\[
\frac{d}{d t}\xi_r = \{\xi_{r},H_{\rm KC}\}= 4 (-H_{\rm KC}) \sqrt{-H_{\rm KC}}/k
\]
Thus $\alpha(H)$ is the frequency of the system which is equal to $\frac{d\,H(J)}{d J}$\,. These results coincide with those given in \cite{hakobyan12,goldstein}. Notice that sometimes the dynamical variables take a different convention:
\[
(J,\xi) \to (J'=2\pi\,J\,,  \xi'= \xi/(2\pi) )
\]

\subsection{Action-angle variables for the HO system}


\begin{itemize}

\item[(1)]
The pair $( p_\phi, \xi_\phi)$. 

This is the same problem as before, for the KC potential, because the angular part is common to all central systems.
Therefore, the first pair of canonical variables is the same,
\begin{equation}
(J_\varphi= p_\varphi \ , \ \xi_\varphi=  \varphi + \arccos(\cot i \cot \theta)\ )
\end{equation}

\item[(2)] The  pair $(\sqrt{L^2}, \xi_\theta)$.

From the expression of the functions $a_\ell^\pm, b_\ell^\pm$ we get
the explicit form of the shift function
\[
\spmc{r}=a_\ell^\pm b_\ell^\mp=\Big(\mp i p_r+\frac{\sqrt{L^2}}{r}+\frac{\omega}{2}r\Big)
\Big(\mp i p_r+\frac{\sqrt{L^2}}{r}-\frac{\omega}{2}r\Big) = \mp 2i\,\sqrt{L^2}\, p_r 
+(-H +2L^2/r^2)
\]
Therefore, $\lambda_\theta$ has the same form  as in KC, while
\[
\sigma_r = \arccos\frac{-H+2L^2/r^2}{\sqrt{H^2 - \omega^2L^2}}
\]
In this case the contribution of $\lambda_\theta$ is two times that of
KC, since the ladder function entering in the symmetry is $(\lpmc{\ell,\theta})^2$:
\[
\lambda_\theta = 2 \arccos(\csc i \,\cos \theta)
\]
We have for the HO,
\[
\{\sqrt{L^2},\Spmc{r,\ell,\theta}\} = \mp 2i \Spmc{r,\ell,\theta}
\]
which implies
\[
\{\sqrt{L^2}/2,\xi_\theta \} = 1
\]
Therefore, the action variable will be $J_\theta= \sqrt{L^2}/2- J_0$
where $J_0$ is a constant. Such a constant is determined from the fact that the action $J_\theta(L^2)$ must be zero for the minimum value of
$L^2$. This minimum value is $L^2= m^2$, (see the expression (\ref{reducedc})), for $\theta= \pi/2$. Then, the action variable with  $J_0 = m/2$ has the form:
\begin{equation}
J_\theta= \sqrt{L^2}/2 - m/2
\end{equation}
In this case, the angle variable is 
\begin{equation}
\xi_\theta =  2\lambda_\theta + \sigma_r  = 
2 \arccos(\sin i \,\cos \theta) +
\arccos\frac{-H+2L^2/r^2}{\sqrt{H^2 - \omega^2L^2}}
\end{equation}
This completes the second canonical pair $(J_\theta,\xi_\theta)$.

\item[(3)]
The pair $(H, \xi_r)$.  

In this case, we dispose only of a pair of ladder functions in order to build the last couple of canonical variables $(J_r(H), \xi_r)$.
From the explicit expression
\[
\lpmc{r}=a^\pm b^\pm=\Big(\mp i p_r+\frac{\ell}{r}+\frac{\omega}{2}r\Big)
\Big(\pm i p_r+\frac{\ell}{r}-\frac{\omega}{2}r\Big) = \mp i\,\frac \omega2 \, p_r 
+(H -\frac{\omega^2}2 r^2)
\]
we get 
\begin{equation}
\xi_r =  \lambda_r    = 
\arccos\frac{H -\frac{\omega^2}2 r^2}{\sqrt{H^2 - \omega^2\ell^2}}
\end{equation}
The PB in this case is
\begin{equation}\label{time}
\{H_{r},\xi_r\} = 2\omega
\end{equation}
and the action has the form:
\[
J_r(H_{r}) = \frac{H_{r}}{2 \omega}- J_0,\qquad \{J_{r},\xi_r\} = 1
\]
where $J_0 = {\ell}/2$.

\end{itemize}

In conclusion, we have obtained via the symmetries, a set of
canonical variables
\begin{equation}\label{aaho}
(p_\varphi, \xi_\varphi;\,  \ell-m, \xi_\theta;\,  \frac{E}{2 \omega}- \frac{\ell}2,\xi_r)
\end{equation}
such that all of them are constants of motion except
$\xi_r$, whose evolution is given by (\ref{time}),
\[
\frac{d}{d t}\xi_r = 2\omega
\]
where the frequency $2\omega$ in this case is constant and it comes from the ladder function of the KC problem.

The expressions of these action-angle for HO and KC systems agree with those given in \cite{hakobyan12}, after
making a correspondence: $(r'=r/\sqrt2,p_r'=p_r\sqrt2, 
{\ell'}^2(\equiv{\cal H}) = \ell^2/2, k'(\equiv \gamma)= k/\sqrt2)$,
where primed variables (and $\gamma,{\cal H}$) are from that reference.

\section{Conclusions and remarks}

Along this paper we have introduced in full detail a method  based on
factorizations to find the symmetries of quantum and 
classical superintegrable systems. We have chosen the KC and HO systems  to prove the consistency and compare this approach with respect to other more conventional treatments.  In the folllowing, we will point out some of the key  points that are specific of this method. Remark that although we have worked out separable systems in spherical coordinates in this paper, it can also be implemented in other types of coordinates. The symmetries obtained are non Hermitian  nor polynomial in the momenta (the polynomial character is generally assumed). We have shown how polynomial symmetries can also be obtained in a simple way within the factorization formalism, allowing us to express our symmetries in terms of the standard ones found in the literature (see (\ref{simho}) for HO or   (\ref{simkc}) for KC). Let us mention these above mentioned points in the quantum and classical contests.

\begin{itemize}
\item Quantum properties.
\\
The symmetries here obtained, in
the quantum formalism,  allow us to find the eigenfunctions of the degeneracy eigenspaces.  Each symmetry operator is made of two types of elementary operators: (i) ladder operators (we use the notation $\lpm{\ell,\theta}$) 
of each of the involutive symmetries  which connect different eigenvalue levels (in this example the eigenvalue $\ell$ of $\hat{L}^2$, see for example (\ref{generating})). In particular, the ladder operators of the Hamiltonian allows to jump among energy levels and they lead to spectrum generating algebras
\cite{manolo12}.
(ii) Shift operators (with the notation $\dpm{r,\ell}$) which change a parameter of a partial Hamiltonian (in this case the parameter $\ell$ in $\hat{H}_\ell(r)$). These shift operators relate (partial) Hamiltonians with different potentials, and in the factorization style are called intertwining operators.

The symmetries obtained in this formalism have non Hermitian form and they include square roots of differential operators (for instance $\ell$, where
$\ell(\ell+1) = L^2$), but
polynomial symmetries in the momenta, are easily recovered. The symmetry algebra can also be determined by using this type of symmetry operators \cite{lissajous14}.

\item  Classical properties.
\\
We have shown how the factorizations can be defined in  classical
systems by means of PBs, and in particular the functions called 
shift and ladder (in correspondence with shift and ladder operators
known from usual factorizations of quantum systems). Such functions do not have the same interpretation as in the quantum case, but they are most useful.
For instance, while ladder operators can be applied in the building
of coherent states \cite{manolo12,cruz08} or to span spectrum generating algebras \cite{kepler12,negro12},   the classical ladder functions give us algebraically the frequency of the motion, or allows to complete the action-angle variables \cite{kepler12}, and also can be used to define a classical spectrum generating algebra.

\item Connection between classical and quantum systems.
\\
The factorization method,  here presented, to find the classical symmetries is
completely new (developed along a list of references), and follows closely the steps of quantum systems.
Some of the most interesting properties are that the symmetries  obtained are  complex functions, whose phase
can be identified in terms of action-angle variables (see (\ref{aakc}) and (\ref{aaho})). This is an
example of the connection of symmetries with Hamilton-Jacobi formalism which was not studied yet in the literature. The orbits (open and closed) are directly obtained from this kind of symmetries, as well as the motion and frequency (corresponding to closed orbits)  by means the ladder functions of the Hamiltonian.

One of the main advantages of this method is that it shows the close connection of  classical and quantum system analysis in a systematic way.
We can follow at each step the quantum and classical expressions and interpret their form. In fact, coherent states is one application where our quantum and classical 
symmetries can be applied in a natural way to establish a bridge for quantum and classical formalisms \cite{manolo12,cruz08}.

\end{itemize}

We have worked out the HO and KC systems in deep detail to make a clear exposition of the basic ingredients and to easily appreciate the differences that one may find with respect to the existing vast literature \cite{hakobyan12,goldstein}.
Our aim in the near future is to showcase the versatility of this method considering its application to other models in quite different contexts such as non-central potential systems that maintain the nest structure \cite{post10,kalnins11,kalnins12,gonera12,post}, parabolic systems \cite{kalnins10b}, systems with spin \cite{ismet}, non Euclidean systems \cite{ballesteros}, applications from exceptional orthogonal polynomials
\cite{ian13,ian15} as well as a dimensional generalization \cite{woj83}.

\section*{Acknowledgments}

This research was supported by Spanish MCIN with funding from European Union NextGenerationEU (PRTRC17.I1) and Consejeria de Educacion from JCyL through QCAYLE project, as well as MCIN project PID2020-113406GB-I00.
\c{S}.~K. thanks Ankara University and the warm hospitality of the Department of Theoretical Physics of the University of Valladolid, where most part of this work has been carried out.

\end{document}